\documentclass[12pt]{iopart}

\usepackage{iopams}
\usepackage{epsfig}

\newcommand{\BE}{\begin{equation}}
\newcommand{\EE}{\end{equation}}
\newcommand{\BA}{\begin{eqnarray}}
\newcommand{\EA}{\end{eqnarray}}

\begin{document}

\title[Birth, death and diffusion of interacting particles]
{Birth, death and diffusion of interacting particles}

\author{Emilio Hern{\'a}ndez-Garc{\'\i}a\ and Crist{\'o}bal L{\'o}pez}

\address{Instituto Mediterr{\'a}neo de Estudios Avanzados, IMEDEA (CSIC -
UIB),\\
Campus Universitat de les Illes Balears, E-07122 Palma de
Mallorca, Spain.}

\ead{emilio@imedea.uib.es}

\begin{abstract}

Individual-based models of chemical or biological dynamics usually
consider individual entities diffusing in  space and performing a
birth-death type dynamics.
In this work we study the properties of a model in this class
where the birth dynamics is mediated by the local, within a given
distance, density of particles. Groups of individuals are formed
in the system  and in  this paper we concentrate on the study of
the properties of these clusters (lifetime, size, and collective
diffusion). In particular, in the limit of the interaction
distance approaching the system size, a unique cluster appears
which helps to understand and characterize the clustering dynamics
of the model.

\vspace*{0.5cm}
{July 6, 2005}

\end{abstract}




\section{Introduction}
\label{sec:intro}

Interacting particle systems are models in which the individual
elements or {\sl particles}, representing molecules, biological
entities, or social agents, are evolved in time following {\sl
microscopic} rules, from which collective macroscopic behavior may
emerge. This modelling approach has been used with success in
physical, chemical, social and biological dynamics
\cite{MarroDickman,Hinrichsen2000,Axelrod,Grimm}. Macroscopic
diffusion in this kind of systems usually arises from the random
walk motion of the individuals.

An important subclass among these models is formed by those in
which birth and death processes occur. Usually they have a
biological inspiration, but they are also found in chemical or
physical contexts. The combination of this number-changing
dynamics with the diffusive motion which is usually assumed for
the particles has profound consequences, being one of the most
striking  the formation of clusters and the apparent attraction
among individuals that are actually noninteracting
\cite{YoungNature,YoungHawaii,Zhang90}. In addition, and most
recently, interactions among individuals that are reflected in
changes in their birth or death rates have also been considered,
and also given rise to a complex collective behavior. In a recent
paper \cite{pre} we explored the consequences on the dynamics of
diffusing individuals induced by the introduction of a finite
spatial extent for the range of such interactions. In particular
we considered the case of a birth rate which becomes reduced
proportionally to the number of particles at a distance smaller
than a range $R$. In a biological setting this dependence models
in a natural way competition for resources, but it can also be a
consequence of other phenomena such as toxin production. The most
notable effect of the interaction was the appearance of a
clustering instability organizing the distribution of individuals
into clusters separated by a typical distance.

 In this paper we address further
characteristics of the model by focussing in such coherent
objects, the clusters: we analyze here their diffusive motion,
their size and some aspects of their dynamics and interactions.
 These are characteristics that lie outside the
capabilities of descriptions in terms of a continuous density of
particles evolving deterministically \cite{pre,physicaD,physicaA},
that have been sometimes used  instead of the ones in terms of
 stochastic particles. The
situation in which the range of interaction is of the order of the
system size, so that any individual interacts with all the others
in the system, allow us to focuss on the dynamics of a single
cluster: Under these conditions a permanent unique cluster emerges
in the system and the studies of the cluster properties are
largely facilitated. It will also be shown that a coarse-grained
deterministic description of the system cannot explain  the
appearance of the cluster in this limit of global interaction of
particles, making emphasis on the importance of the fluctuations
(discrete nature of the particles) in the model. Although the
model was originally introduced in two spatial dimensions, we
restrict here to the onedimensional case, since it contains the
essentials of the phenomenology.

The paper is organized as follows: In the next section the model
is presented. In Section 3 we explain the clustering instability
of the system through the introduction of a spatiotemporal field
related to a net growth rate. In section 4 we study the properties
of the system in the limit of global coupling, focussing in the
dynamics of cluster competition, cluster size and diffusion
properties of the cluster. Section 5 presents our conclusions.

\section{A model of interacting random walkers}
\label{sec:model}

Ensembles of particles performing Brownian motion and with a
birth-death type dynamics has long been used in the modelling of
biological populations. A natural way of introducing an effective
interaction among the organisms when there is a competition for
the sources is to consider that birth and death rates of any
individual are altered by the local density in its neighborhood.
With this in mind, the authors recently introduced a model where
the birth rate for a given particle is decreased with the number
of other particles that are within a finite distance $R$. More in
detail, the system consists initially in a set of $N$ particles
randomly located at positions $x_1,x_2, ..., x_N$ in a
onedimensional segment of length $L$ with periodic boundary
conditions. The number $N$ and the positions of the particles are
evolved according to the following algorithm: First, one of the
particles is chosen at random (let us call it the $i$ particle, at
position $x_i$). Second, the basic ingredients of the model, the
probabilities $\lambda_i$ and $\beta_i$ of reproduction and death
respectively, which depend on the environment surrounding $i$, are
calculated as described later. Third, with the birth probability
$\lambda_i$ a new particle is introduced in the system exactly at
the location of the mother particle $i$, or rather, with the death
probability $\beta_i$, the particle $i$ disappears from the
system. With probability $r_i=1-\lambda_i-\beta_i$, no changes are
made. These three steps leading to the trial of a particle and of
its fate are repeated a number $N$ of times, after which the
number of particles $N$ in the system is updated. We choose this
lapse of $N$ trials to be the unit of time, so that $\lambda_i$
and $\beta_i$ are also, at least at the beginning of each time
unit, birth and death probabilities per particle and per unit of
time. After this, each particle is moved a random distance drawn
from a Gaussian distribution of variance $\sigma^2$. This Brownian
motion leads to macroscopic diffusion with diffusion coefficient
$D=\sigma^2/2$. Then the process repeats for the following time
units.

The defining characteristic of the model is that the birth and
death probabilities $\lambda_i$ and $\beta_i$ may depend on the
number of other particles $N_R^{i}$ within a distance $R$ of the
chosen one $i$ (i.e. within an interval of size $2R$ centered on
particle $i$). To compare with previous works we include this
dependence only on the birth rate:
\BA
\lambda_i &=& \max \left(\lambda_0-\frac{N_R^{i}}{N_s},0 \right)
\label{birth} \\
\beta_i &=& \beta_0
\label{death}
\EA
The $\max$ function is introduced to avoid negative values for the
probability. $N_s$ is a saturation constant. In addition we choose
$\lambda_0+\beta_0=1$, so that the independent model parameters
are $D,R,N_s$ and $\beta_0$ (and the system size $L$). Instead of
$\beta_0$ one can characterize the system by the maximum growth
rate $\mu_0 \equiv \lambda_0-\beta_0 = 1-2 \beta_0$.

\section{The clustering instability}
\label{sec:clustering}

\begin{center}
\begin{figure}
\epsfig{file=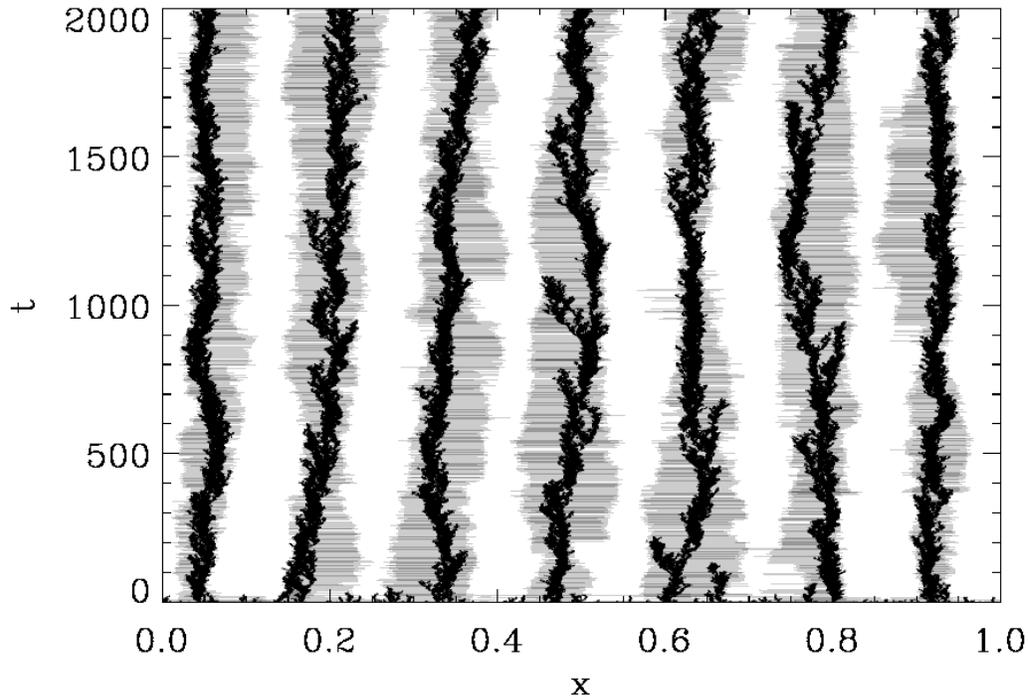,width=0.9\linewidth} \caption{
Spatiotemporal representation (space on the horizontal axis, and
time on the vertical)
 of the patterns of our stochastic
model. $R=0.1$, $N_s=50$, $\lambda_0=0.8$, $\beta_0=0.2$,
$D=5\times 10^{-7}$. Positions of the particles are plotted in
black. They cluster in seven groups during most of the simulation.
The background gray level field displays the local net growth
$\mu(x,t)$. White denotes the most negative value
$\mu(x,t)=-\beta_0$, whereas the darkest shades of gray denote
values fluctuating around zero.}
\label{fig:particlepattern}
\end{figure}
\end{center}

A simple mean-field argument would predict that, assuming a steady
homogeneous density of particles $\rho_0$, the expected number of
particles at distance smaller than $R$ from a given one would be
$N_R^{i} \approx 2R\rho_0$, and thus the effective birth rate will
be given by $\lambda_0-2R\rho_0/N_s$. In a statistically steady
state this should equate the death rate $\beta_0$, so that $\rho_0
\approx \mu_0 N_s/2R$. A typical evolution of the system when the
diffusion coefficient is small and $\mu_0$ positive and not too
close to zero is shown in \fref{fig:particlepattern}. The salient
feature is the grouping of the particles in a number of
fluctuating clusters. The first hypothesis in the mean field
argument, namely the existence of a homogeneous density, is
clearly inappropriate. In fact the total number of particles is
larger than the $\rho_0 L$ predicted by the argument. One can
understand the observed pattern from the following reasoning: If,
after starting with a random distribution of particles, there is a
fluctuation in the particle positions such that the local density
increases at two relatively narrow locations (the cluster seeds)
separated by a distance between $R$ and $2R$, particles left close
to the middle point between these locations will count the
population of the two clusters among their neighbors, whereas
particles in each cluster will count only as neighbors the
particles in the own cluster. As a consequence the birth rate in
the cluster seeds will be larger than the one in the region in
between, and the difference between particle density in both
regions will increase leading to an instability that will finally
concentrate all the particles in big clusters at a separation
$fR$, intermediate between $R$ and $2R$. In
\fref{fig:particlepattern} the separation is $fR$ with $f\approx
1.4$. Within this reasoning, all particles in a cluster feel
essentially the same $\lambda_i$ and $\beta_i$, and do not
interact with particles in the other clusters if the cluster
positions do not approach too much. Then the number of particles
in each cluster $N_c$ will stop growing when it reaches a level
such that $\lambda_i \approx \beta_i$, i.e. $N_c \approx \mu N_s$.
This gives $N_c \approx 30$ for each cluster in
\fref{fig:particlepattern}, whereas in the simulation this number
fluctuates around $29$. It is a curious and counterintuitive
effect that, since $f<2$, the mean number of particles in the
system $\mu_0 N_s L/fR$ is larger than in the homogeneous
situation $\mu_0 N_s L/2R$, despite the fact that grouping
particles into clusters seems at first sight to increase $N_R^{i}$
and thus to decrease the birth rate.

To give further support to the arguments above we plot in
\fref{fig:particlepattern} a net growth rate field $\mu(x,t)$
defined as
\BE
\mu(x,t)=\max \left(  -\beta_0 ,
\lambda_0-\beta_0-\frac{1}{N_s}\int_{x-R}^{x+R} \rmd u \hat
\rho(u,t) \right)
\label{muxt}
\EE
where $\hat \rho(x,t) \equiv \sum_{k=1}^{N(t)} \delta\left(
x-x_k(t)\right)$ is the microscopic density of particles. The
integral in this expression counts the number of particles in the
neighborhood of a point $x$, so that the field $\mu(x,t)$ would
give essentially the net particle number growth rate when
evaluated at the particle positions ($\mu(x=x_i,t) \approx
\lambda_i-\beta_i$) except for the extra counting of the particle
$i$ which is negligible in situations with many particles. The
integral acts as a low-pass filter of the microscopic density, so
that typical features in $\mu(x,t)$ will have sizes of order $R$
and above. \Fref{fig:particlepattern} displays this field and
confirms that its value is $-\beta_0$ in most of the intercluster
region, so that only death can occur there, and that it is slowly
varying spatially (and temporally fluctuating close to zero) in
regions of size close to $R$, which we call the {\sl niches},
around the places where the particles are. The particles are in
fact so concentrated inside the niche that there are no important
differences among the values of $\mu(x,t)$ they feel at most of
the times $t$. The introduction of the field $\mu(x,t)$ is a
convenient tool to represent the interactions at a distance
between the particles in terms of the local interaction of each
particle with the field $\mu(x,t)$ at its location.

The above heuristic calculations fail at least in two situations.
First, they are qualitative arguments involving average
quantities, thus they will give wrong results when statistical
fluctuations are strong. This happens when the number of particles
in the clusters is small, which occurs when decreasing $\mu_0$. It
was shown in \cite{pre,physicaD,physicaA} that fluctuations lead
to full extinction for $\mu_0$ below a $\mu_c$ that for $R,N_s$,
and $D$ as in \fref{fig:particlepattern} is given by $\mu_c
\approx 0.34$, which is is much above the value $\mu_c \approx 0$
that one would estimate if  fluctuations were neglected. This is
so because the state with zero particles is an {\sl absorbing
state} \cite{mamunoz} from which no recovery is possible. This
irreversibility biases the statistics so that extinction occurs as
soon as the average particle number becomes of the order of its
fluctuations, despite that they act both increasing and decreasing
the number of particles. Close to $\mu_c$ the expected number of
particles in any cluster is smaller than implied by the above
estimations, in the same way as shown in \cite{physicaD,physicaA}
for expected densities. This absorbing character of the empty
state has also the consequence that a intercluster spacing between
$R$ and $2R$ is the expected outcome only from an initial
condition of particles filling up the whole system. Initial
conditions in which particles are already clustered into groups
more distant than $2R$ will also be stable, despite of the fact
that a niche will develop in the empty space in between. This is
so because of the impossibility of spontaneous creation of
particles in that empty niche, and of the difficulties for
particles in neighboring niches to colonize the empty one due to
the presence of {\sl zones of death} around it, and of the strong
correlations in the motion of all the particles inside a cluster
that will be discussed in \sref{subsec:diffusion}.

A more subtle failure occurs even at sufficiently large $\mu$. The
arguments above rely on the assumption that the clusters remain
sufficiently narrower than $R$. In fact this is what happens in
the situation of \fref{fig:particlepattern} and always for
sufficiently small $D$. But the mechanism that keeps the cluster
at this width without spreading over all the niche is by no means
obvious, and will be discussed in section \ref{subsec:cluster}. It
is natural to think, and it will be confirmed later, that
increasing the diffusion coefficient $D$, a quantity that has not
yet appeared in our argumentations, will increase the cluster
size. For sufficiently large $D$ configurations will become
homogenized and clustering will not be observed.

Another hypothesis in our reasoning, confirmed by the simulations,
is that there is a single cluster in each niche. Therefore, it
seems  that it is essential for the understanding of the full
dynamics  to understand the behavior -- size, motion, stability,
... -- of single clusters. We do that in the following section,
but after noting that the simplest situation one can consider is
that in which the range of interaction $R$ reaches the full system
size (i.e., $R=L/2$) since then there is a single niche in the
system. This case should represent the dynamics of particles
inside any of the niches (and thus being noninteracting with the
other clusters) in most of the situations, except when the cluster
approaches the niche boundaries, that would produce an enhanced
mortality at the exposed side.

\section{Dynamics with global coupling}
\label{sec:global}

The case $R=L/2$ is the simpler one since then all the birth rates
$\lambda_i$  become completely independent of the position $x_i$
of the particles, and $\mu(x,t)$ fluctuates only in time, not in
space, adapting to the total number of particles in the system:
$\mu(x,t)=\mu(t)=\mu_0-N(t)/N_s$, if $N(t)<\lambda_0 N_s$, and
$\mu(t)=-\beta_0$ for larger $N(t)$. We expect $N(t)$ to fluctuate
around $N_c=\mu_0N_s$. Next, we first explain why there is still a
single cluster in the system, despite the large extent of the
niche, and then estimate its width and diffusive behavior.

\subsection{Cluster competition and lifetime}
\label{subsec:competition}

In principle, since in this limit $R \to L/2$ the interactions among
 the particles are
independent of their positions, and if we forget about stochastic
effects, the system can be organized in a single cluster with
$N_c$ particles, or in two with $N_c/2$ particles each, or in fact
in any configuration such that the total number of particles
remains in average close to $N_c = \mu_0 N_s$. But not all these
configurations are equally stable against stochastic particle
number fluctuations. Clusters with a small number of particles
will disappear quite fast because of the high probability of a
number fluctuation of the order of its size, and the
irreversibility of  cluster extinction. This effect will be
further on facilitated by the negative correlations between particle
number fluctuations of different clusters (a positive fluctuation
in one cluster will reduce $\lambda_i$ for all particles, so that
birth will be relatively less frequent in the following steps). We
expect that multiple cluster configurations will decay into the
most stable state, the one consisting of a single cluster that
will be the one observed for most of the time, independently of
the extension of the niche.

\begin{center}
\begin{figure}
\epsfig{file=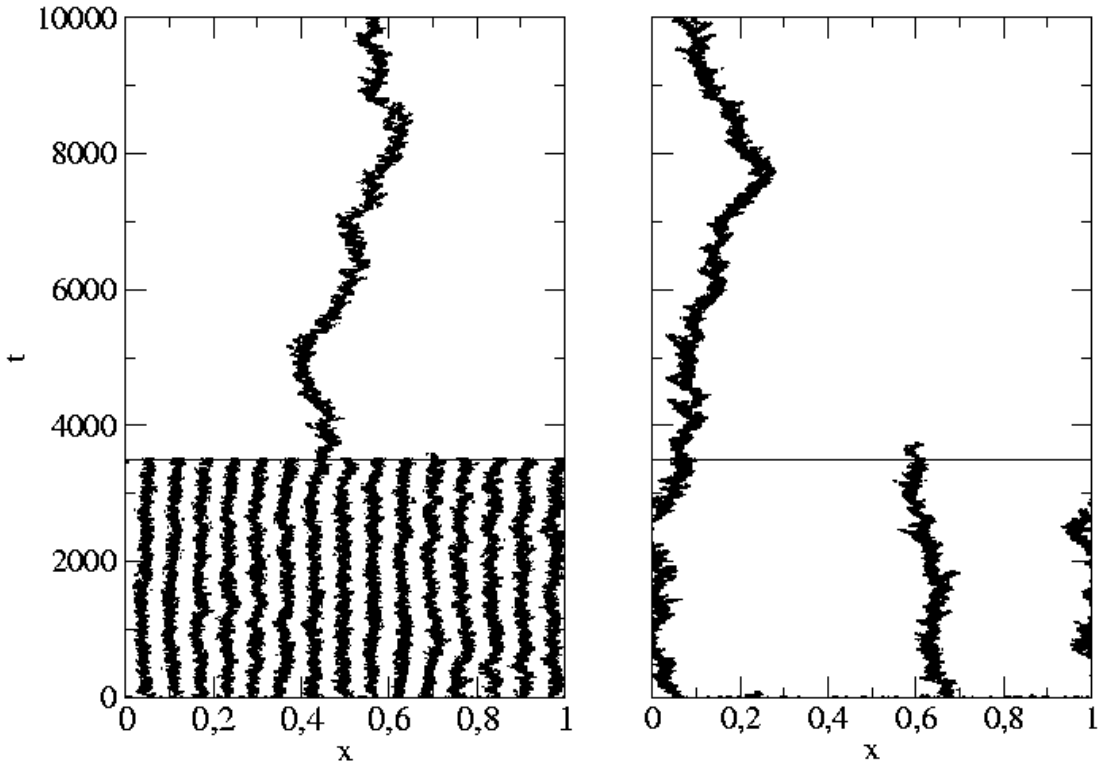,width=0.8\linewidth, angle=-90}
\caption{ Left panel: evolution of a set of of $15$ clusters
prepared under $R=0.05$ before we change $R$ to global coupling
$R=L/2$ at time $t_i=3500$ (indicated by the horizontal line). A
short time afterwards, a single cluster survives. Right panel:
analogous simulation but with only two clusters (prepared under
$R=0.324$) before switching to global coupling at $t_i=3500$. The
rest of parameters take the values $\beta_0=0.25$, $D=10^{-6}$,
and $N_s=50$. }
\label{fig:competition}
\end{figure}
\end{center}

\begin{center}
\begin{figure}
\epsfig{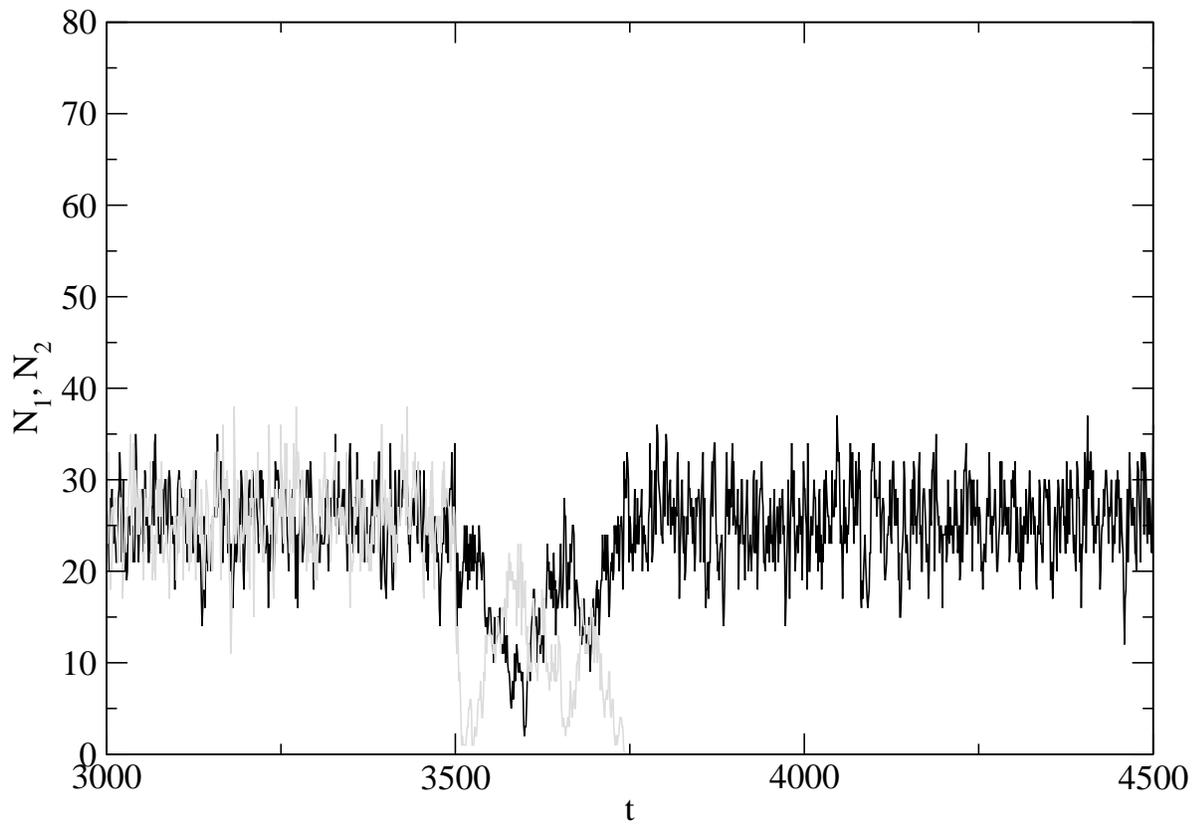}
\caption{Time evolution of the number of particles of the two
clusters in the right panel of \fref{fig:competition} just before
and after the time ($t_i=3500$) at which $R$ is switched to global
coupling. In black, data for the surviving one, in gray, data for
the one disappearing at $t \approx 3750$.}
\label{fig:numbers2clusters}
\end{figure}
\end{center}

\Fref{fig:competition} shows the evolution of initial conditions
containing several clusters under global coupling. The initial
part of the evolution was generated with small values of $R$ so
that the system organizes in a number of clusters separated by
distances $fR$. $R$ is increased to $R=L/2$ at $t_i=3500$. As
expected, all the clusters except one disappear in a very short
time. \Fref{fig:numbers2clusters} shows the evolution of the
number of particles in each cluster in the case of two competing
ones. Immediately after the change in $R$ the total number of
particles in the system decreases, since the initial $N$ is twice
what it should be at the end, $N_c$. Once the total number of
particles is close to $N_c$ the real competition starts. One of
the clusters becomes smaller by chance and disappears after some
time, victim of stronger relative fluctuations. Then, the survivor
cluster adjusts its population to fluctuate close to the expected
final value $N_c=\mu_0 N_s$ (which in the figure
\Fref{fig:numbers2clusters} is $N_c =25$ because $\mu_0=0.25$ and
$N_s=50$). The tendency of the single cluster to maintain this
number of particles will make it much longer lived. On general
grounds we expect the time of extinction of a single cluster to
increase exponentially with its number of particles $N_c$ so that
for practical purposes single clusters remain forever when they
have more than a few particles.

We can estimate the fluctuations in the particle numbers of
competing clusters and their lifetimes by adapting some arguments
from \cite{Zhang90}. We can compare with the models in that paper
if we introduce two approximations: First, that the total number
of particles is exactly $N=N_c=\mu_0 N_s$ instead of fluctuating
around this number. Second, we perform the calculations by
thinking that all the $N$ particles are checked for reproduction
or death exactly once every unit of time, instead of being checked
once {\sl on average}. With these approximations, the number of
particles in one of the clusters, say cluster $1$, is given by
\BE
N_1(t)=N_1(t=0)+\sum_{\tau=1}^t \Delta N_1(\tau)
\label{N1}
\EE
where $\Delta N_1(\tau)$ is the increment in particle number in
that cluster occurring during step $\tau$. This quantity is given
by
\BE
\Delta N_1(\tau)= \sum_{i=1}^{N_1(\tau)} \alpha_i(\tau)
\label{DeltaN1}
\EE
$\alpha_i(\tau)$ is the increment or decrease in particle number
occurring when particle $i$ is checked at time $\tau$, i.e. $+1$
with probability $\lambda_0-N(\tau)/N_s \approx \beta_0$, $-1$
with probability $\beta_0$, and $0$ with probability $1-2\beta_0$.
We have used the assumption that the total number of particles in
all the clusters $N(\tau)$ has a constant and nonfluctuating
value: $\mu_0 N_s$, and despite this constraint, we still assume
that the $\{\alpha_i(\tau)\}$ are independent variables for each
$i$ and $\tau$. These approximations will be more consistent when
$N_1 \ll N$. {}From them we get
\BE
\left< \alpha_i(\tau)\right> = 0  \ \ ,  \ \  \left< \alpha_i(\tau)
\alpha_j(\tau')\right> = \delta_{ij}\delta_{\tau \tau'} \left<
\alpha_i(\tau)^2\right>  \ \ ,  \ \  \left<
\alpha_i(\tau)^2\right>=2\beta_0
\EE
{}from which $\left< N_1(\tau)\right> = N_1(0)$ and
\BE
\fl \left< \Delta N_1(\tau)\right> = 0  \ \ ,  \ \ \left< \Delta
N_1(\tau) \Delta N_1(\tau')\right> = \delta_{\tau \tau'} \left<
\Delta N_1(\tau)^2\right> \ \ ,  \ \   \left< \Delta
N_1(\tau)^2\right>=2\beta_0 N_1(0)
\label{averageincrements}
\EE
Thus we see that the expected number of particles in each cluster
is a constant, but the variance increases without limit as follows from
(\ref{N1}) and (\ref{averageincrements}):
\BE
\left< N_1(t)^2\right> - \left<N_1(t)\right>^2 \approx 2 \beta_0 t
N_1(0)
\label{varianceN1}
\EE
A natural identification of the time of extinction of the cluster
is the time $t_m$ for which the variance equates the square of the
mean value of $N_1$. This happens when
\BE
t_m \approx \frac{N_1(0)}{2 \beta_0}
\label{tm}
\EE
Clearly, this expression can not be applied to the lifetime of a
single cluster, since then one can not neglect the fluctuations in
$N$, nor the correlations among $N$ and the reproduction rates.

\begin{center}
\begin{figure}
\epsfig{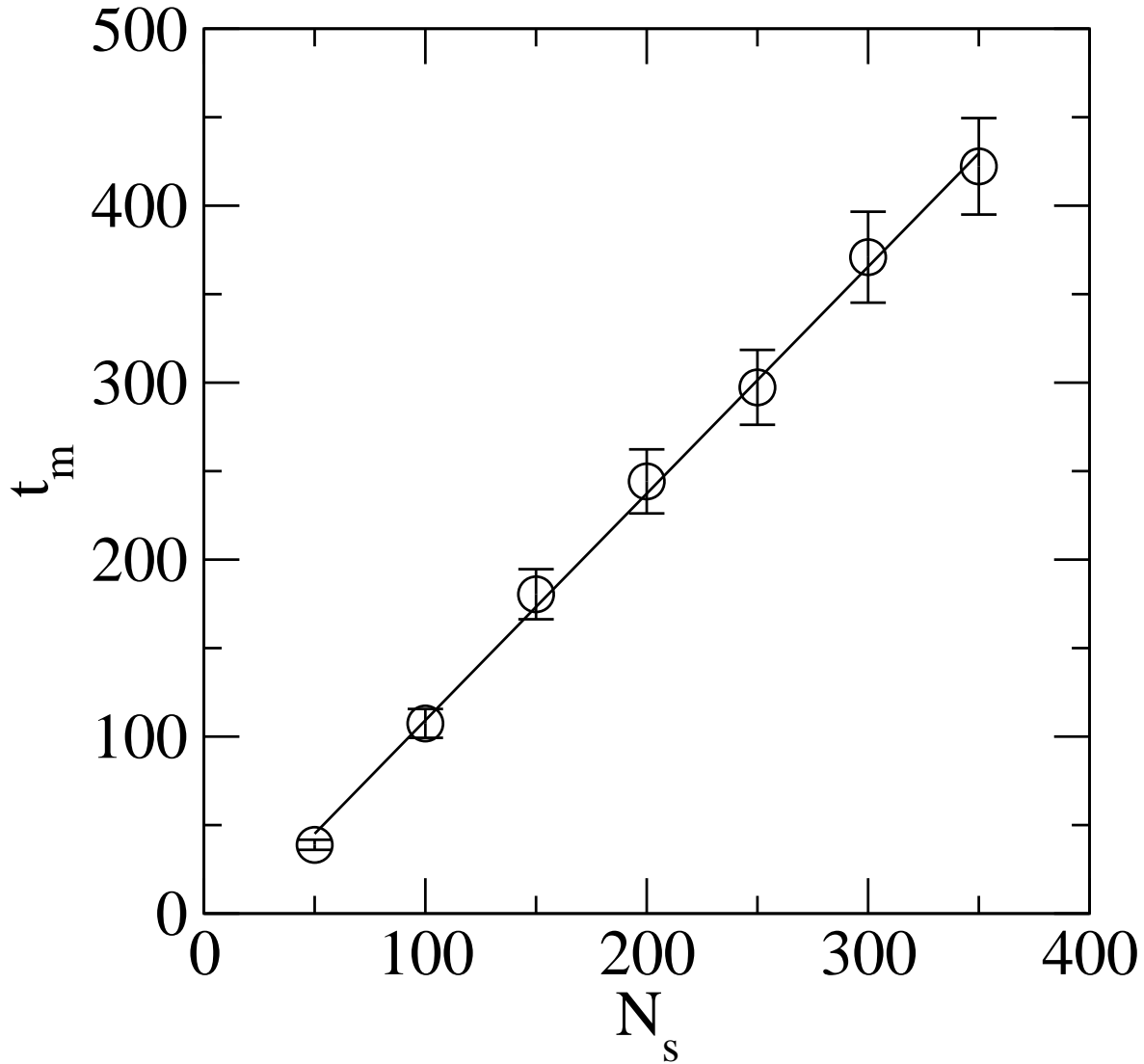} \caption{The average time
(over $500$ realizations) as a function of $N_s$ for extinction of
one of the two competing clusters evolving under global coupling.
$D=10^{-5}$ and  $\beta_0=0.25$. The initial number of particles
in the cluster $N_1(0)$ is approximately equal to $\mu_0 N_s$. The
line is the best linear fit to the data, with slope $1.2$. }
\label{fig:tiempos}
\end{figure}
\end{center}

\Fref{fig:tiempos} shows the average time for disappearance of one
of the two clusters evolving in situations such as the right panel
of \fref{fig:competition} or \fref{fig:numbers2clusters}. We plot
this time versus $N_s$ (the controllable parameter of the model)
which approximates $N_1(0)$ via $N_1(0) \approx \mu_0 N_s$. In the
figure, $\beta_0=0.25$ so that the slope of the curve should be
$1$. The slope of a linear fitting through the numerical data is
$1.2$. It is remarkable that despite the severe approximations
introduced to obtain (\ref{tm}) the general trend is correct and
quantitatively close to the numerically observed lifetimes.

\subsection{Cluster spatial size}
\label{subsec:cluster}

Once we have justified that a single cluster in a niche is the
natural state of the evolving particle system, we can now estimate
the spatial extent of this cluster. Since particles do not attract
nor repel, they will experience pure diffusive motion during all
its lifetime. The only factor impeding the unbounded diffusive
spreading of the cluster is the fact that the lifetime is finite.
One is tempted to guess that the size of the cluster would be
$\sqrt{2D/\beta_0}$, since this is the diffusive displacement of a
particle during its average lifetime $\beta_0^{-1}$ (we define
cluster size as the standard deviation of the particle positions:
$S_c\equiv \left( <x_i^2>-<x_i>^2 \right)^{1/2}$; an alternative
definition as the root mean square of the distance between two
randomly chosen particles is larger by an extra factor
$\sqrt{2}$). That guess of the size is incorrect because the
cluster width will continue to grow after the initially chosen
particle has died if its descendants (its {\sl family}) are still
alive and diffusing (we can say that the descendants continue the
diffusion process of the mothers, since they are born at the
mothers location). The width of the cluster will be determined
from the diffusive spread among the members of the longest lived
families. We can not apply the arguments for lifetimes of the
previous subsection to a whole cluster, but they can be applied to
subsets inside a cluster (in the case of global coupling there is
no difference between the interactions among particles in
different or in the same cluster) as for example a family. The
longest lived families would be the ones reaching a size close to
the total one $N_c$. From the estimation of the previous section,
the typical lifetime of a family is of the order of $t_m=\mu_0
N_s/2\beta_0$. The standard deviation reached by the particle
positions after that time will be $\sqrt{2Dt_m}$. The typical
cluster size is thus given by
\BE
S_c \approx \sqrt{\frac{DN_s(1-2\beta_0)}{\beta_0}}
\label{Sc}
\EE

\begin{center}
\begin{figure}
\epsfig{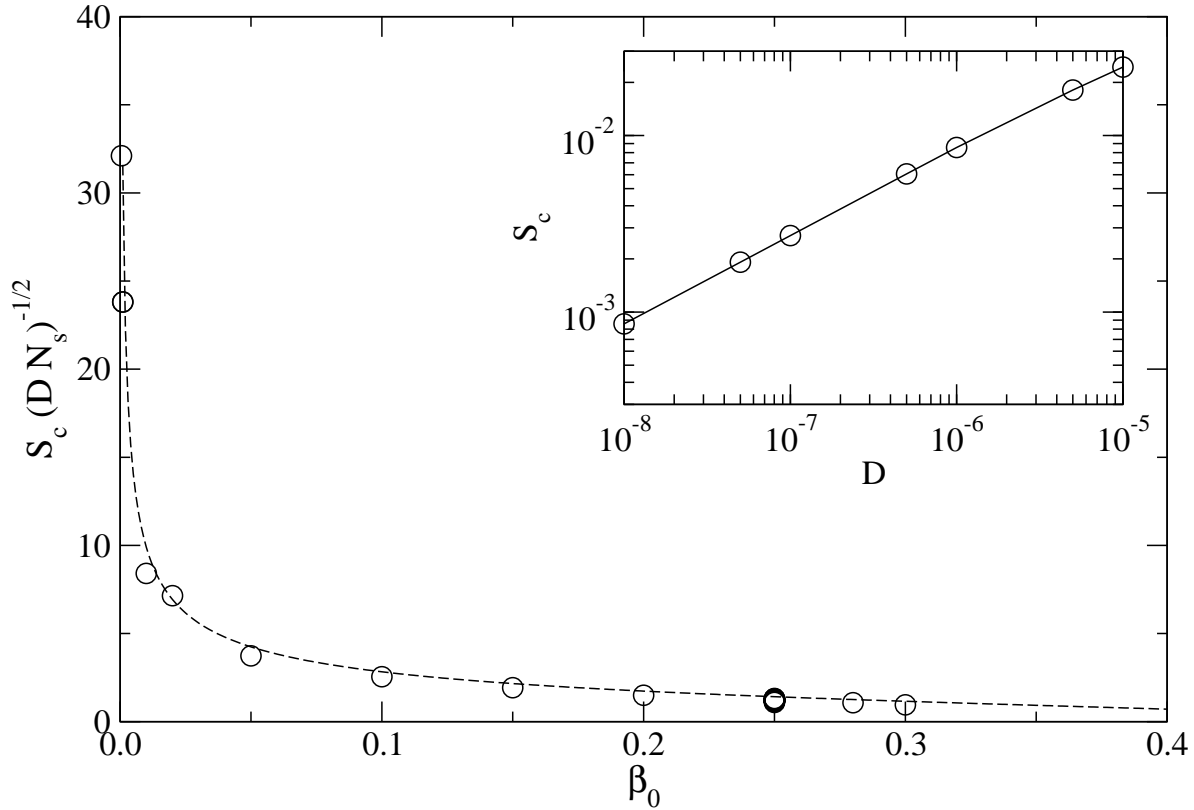} \caption{Cluster
size, estimated as the mean square dispersion of the particles,
and averaged in time, as a function of $\beta_0$. The parameter
values are $D=10^{-7}$, $N_s=50$ and $R=0.5$ (global coupling).
 The vertical axis
is scaled with $\sqrt{D N_s}$, and the solid line is the function
$\sqrt{(1-2\beta_0)/\beta_0}$ predicted by (\ref{Sc}). For
$\beta_0=0.25$, and always $R=0.5$, other values of the parameters
has also been plotted (collapsing almost in the same point):
$D=10^{-7}$ and $N_s= 40, 50, 60, 70, 80, 90, 100, 150, 200, 250,
300$; $N_s= 50$ and $D=10^{-9}, 10^{-8}, 10^{-6}, 10^{-5}$.
 The inset, for $\beta_0=0.25$ and $N_s= 50$, shows the clear diffusive dependence $S_c
\sim D^{1/2}$ of the cluster size.}
\label{fig:sizes}
\end{figure}
\end{center}

\Fref{fig:sizes} shows that this expression, obtained under
several approximations, is surprisingly accurate.

\subsection{Cluster diffusion}
\label{subsec:diffusion}

In figures \ref{fig:particlepattern} or \ref{fig:competition} we
see that each cluster as a whole undergoes a kind of random walk
(in \fref{fig:particlepattern} only until the clusters touch the
limit of the niche, moment at which some particles interact with
the neighboring niche, some of them die, and the remaining ones
are the ones returning to the interior of the niche). If naively
one forgets the correlations among the diffusing particles one
would guess that the diffusion coefficient (estimated from the
displacement of the center of mass) should be $D/N_c$.
\Fref{fig:diffusion} shows that in fact the diffusion coefficient
is $D$, the same as the one for the individual particles. The
particles in the cluster move so coherently that they behave
collectively as a single particle. We can understand this result
again by modifying a discussion from \cite{Zhang90}:

\begin{center}
\begin{figure}
\epsfig{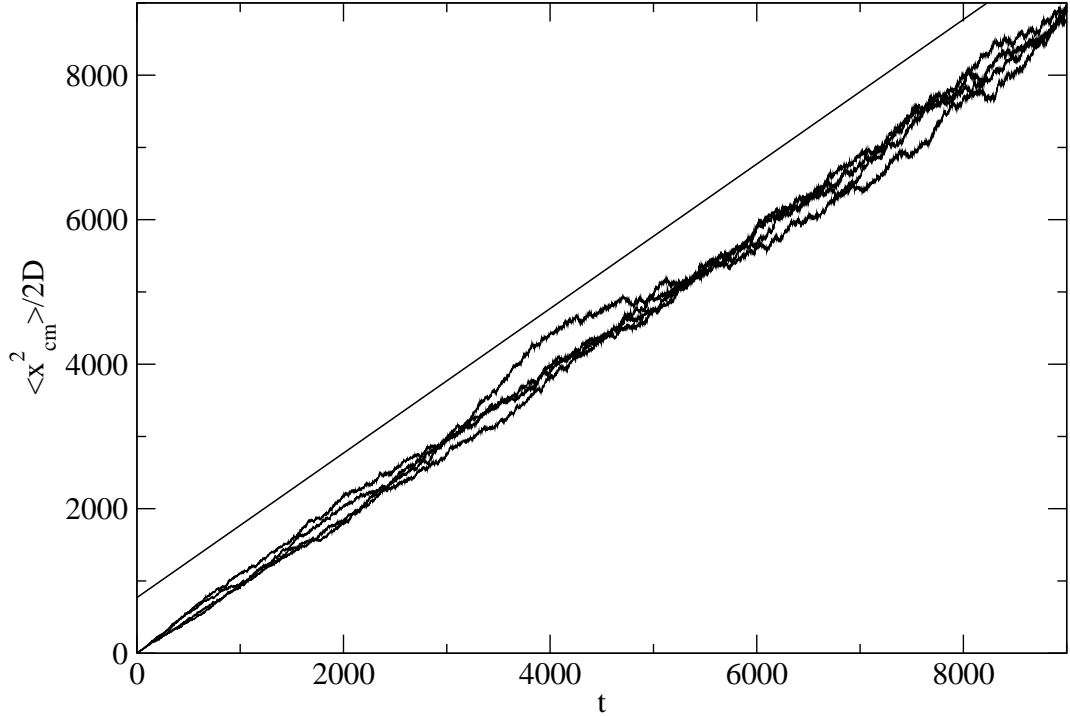} \caption{ Position
(average over $200$ realizations) of the center of mass of one
cluster, $<X_{cm}^2>$, scaled with $2 D$, vs time, and for
different values of $D$, $\beta_0$ and $N_s$. The slope of the
straight line is $1$ confirming that the center of mass diffuses
with coefficient $D$. Four different curves are plotted, that
almost collapse on a single one. They correspond to the following
parameter sets: a)$D=10^{-6}$, $\beta_0=0.20$, $N_s=50$; b)
$D=10^{-6}$, $\beta_0=0.25$, $N_s=80$; c) $D=10^{-7}$,
$\beta_0=0.25$, $N_s=50$; d) $D=10^{-5}$, $\beta_0=0.25$,
$N_s=50$. }
\label{fig:diffusion}
\end{figure}
\end{center}

We begin by splitting the expression for the center of mass
$X_{cm}$ at time $t-1$ as follows:
\BA
\fl X_{cm}(t-1) \equiv \frac{1}{N}\sum_{i=1}^N x_i(t-1) \nonumber  \\
\lo= \frac{1}{N} \sum_{i \in F} x_i(t-1) + \frac{1}{N} \sum_{i \in
B} x_i(t-1) +\frac{1}{N} \sum_{i \in M} x_i(t-1).
\label{xcmt-1}
\EA
$B$ is the set of particles that will give births during the next
time step, $M$ is the set of particles that will die, and $F$ the
remaining ones. As before, we assume that the number of particles
$N$ is constant and equal to $N_c=\mu_0 N_s$. After one time step
the birth, death, and diffusion process will change the above
expression into:
\BA
\fl X_{cm}(t) \equiv \frac{1}{N}\sum_{i=1}^N x_i(t-1)= \frac{1}{N}
\sum_{i \in F} \left( x_i(t-1) +
\sigma g^F_i(t)\right) \nonumber  \\
\lo+\frac{1}{N} \sum_{i \in B} \left( x_i(t-1) + \sigma
g^B_i(t)\right) +\frac{1}{N} \sum_{i \in B} \left( x_i(t-1) + \sigma
h^B_i(t)\right)
\label{xcmt}
\EA
The particles in $M$ have disappeared, and there are two copies of
the particles in $B$. $g^F_i(t)$, $g^B_i(t)$ and $h^B_i(t)$ are normalized and
independent Gaussian random numbers, and $\sigma=\sqrt{2D}$.
Substraction of (\ref{xcmt}) and (\ref{xcmt-1}) gives (remember that
we are assuming a constant total number of particles)
\BE
X_{cm}(t)-X_{cm}(t-1)=\frac{1}{N} \sum_{i \in B} x_i(t-1) -
\frac{1}{N} \sum_{i \in M} x_i(t-1) + \frac{\sigma}{\sqrt{N}} g(t)
\label{xcmincrement}
\EE
We see that, in addition to the standard diffusive motion given by
the term containing the normalized Gaussian number $g(t)$, the
displacement of the center of mass is also controlled by the
removal of the particles in $M$ and the replication of particles
in $B$. To explicitly estimate this contribution we approximate
the number of particles in $B$ and $M$ by its expected value
$n\equiv N\beta_0$. Thus the first two terms in the right hand
side of (\ref{xcmincrement}) can be written, after renaming the
particle labels, as
\BE
\frac{1}{N}\sum_{i=1}^{n} \left( x_i(t-1)-x_{i+n}(t-1)\right)
\EE
The term between parenthesis is the distance between two randomly
selected particles inside the cluster. Its typical (root mean
square) value is $\sqrt{2}S_c$. If $n$ is large enough we can
invoke the law of the large numbers and find
\BE
\left< \left( X_{cm}(t)-X_{cm}(t-1) \right)^2\right> \approx
\frac{2D}{N} + \frac{n}{N^2}2 S_c^2 \approx \frac{2D}{N} + 2D
\label{xcrms}
\EE
In large clusters the first term, coming directly from the
uncorrelated particle motion, becomes negligible, and the second,
coming from the reproductive correlations, i.e. correlations
introduced in the system when new particles are born at the
mothers positions, dominates. This demonstrates that the diffusion
coefficient of big clusters approaches $D$, the diffusion
coefficient of a single particle. Reproductive correlations
transform individual random walks into fully coherent motion of
the particles inside a diffusing cluster.

\section{Conclusions}

We have described results for the behavior of an interacting
particle system evolving under diffusion, birth and death
processes. Because of its simplicity, and of the ubiquity of the
interactions considered (essentially competition for resources) we
expect this behavior to represent features of real systems, in
particular biological ones. The essential building blocks of the
dynamics are particle clusters that spontaneously form in the
parameter range considered here. They live in niches where birth
and death rates equilibrate. Previous work \cite{pre} described
the appearance and spacing of the clusters in terms of a
deterministic equation for a continuous density, but it failed to
reproduce many important properties of the model. Here we have
noticed the adequacy of the global interaction limit to represent
the dynamics inside a single niche, and thus concentrate in that
limit to understand features not well described by the
deterministic continuous description. It is worth mentioning that
a deterministic continuous description is particularly inaccurate
in this limit: the integrodifferential equation in \cite{pre} for
an expected density of particles $\rho(x,t)$ becomes
\BE
\partial_t\rho(x,t)=\left(
\mu_0-\frac{N(t)}{N_s}\right)\rho(x,t)+D\partial_x^2\rho
\label{meanfield}
\EE
where the dynamics of the total number of particles is decoupled
from the spatial distribution:
\BE
\dot N(t) = \mu_0 N(t)-\frac{1}{N_s}N(t)^2
\label{logistic}
\EE
Equations (\ref{meanfield}) and (\ref{logistic}) can be solved
exactly. At long times $N(t) \rightarrow \mu_0 N_s$, in agreement
with the calculations in this paper, but then \eref{meanfield}
becomes a diffusion equation, predicting a homogeneous particle
distribution al long times. This is in contrast with the grouping
of all the particles in a single cluster that is observed in
simulations of the stochastic particle model and understood by the
previous calculations. The situation is very similar to the one
described in \cite{YoungNature}. Reproductive correlations, the
phenomena produced by the fact that newborns appear at the mother
location whereas death can occur anywhere, are the responsible for
shaping the clusters and their motions, despite that a continuous
equation that neglects them may give the correct spacing. It was
emphasized in \cite{YoungNature} that an equation for an expected
density is just the first step in the hierarchy describing
particle distributions and that additional quantities --
correlation functions -- are needed to fully capture the
clustering behavior. In a remarkable recent paper \cite{Birch},
Birch and Young describe that hierarchy in general terms and they
are even able to solve it in the case of global coupling for a
model of interactions in which the death rate is neighborhood
dependent, and for which the death probability vanishes when there
is a single particle in the system, so that complete extinction is
impossible. It would be very interesting to understand the
relationship between the results for their model and the ones
presented here. Without using the power of that general framework
we have obtained here expressions for lifetimes, sizes and
collective diffusion that agree very well with numerical
simulations.

\ack

Financial support from MEC (Spain) and FEDER through project
CONOCE2 (FIS2004-00953) is greatly acknowledged. C.L. is a {\sl
Ram{\'o}n y Cajal} fellow.

\end{document}